# A VARIATIONAL DEDUCTION OF SECOND GRADIENT POROELASTICITY II: AN APPLICATION TO THE CONSOLIDATION PROBLEM

ANGELA MADEO, FRANCESCO DELL'ISOLA, NICOLETTA IANIRO AND GIULIO SCIARRA

The second gradient model of poromechanics, introduced in Part I, is here linearized in the neighborhood of a prestressed reference configuration to be applied to the one-dimensional consolidation problem originally considered by Terzaghi and Biot. Second gradient models allow for the description of boundary layer effects both in the vicinity of the external surface and the impermeable wall.

The formulated differential problem involves linear pencils of ordinary differential operators on a finite interval, with boundary conditions depending on the spectral parameter. Taking into account the dependence of the differential problem on initial stresses a linear stability analysis is carried out. Finally, numerical solutions are compared with the corresponding classical Terzaghi solutions.

## 1. Introduction

This paper addresses a geotechnical application of the macroscopic second gradient poroelasticity theory presented in the first part; in particular we aim to treat the well known soil consolidation problem [Terzaghi 1943]. The consolidation of a soil layer of depth $L$ can be schematically described as follows: when an external load $p^{\text{ext.}}$ is applied on the surface of the layer, the fluid starts moving from the layer towards the surface, and it finally leaves the system. While the fluid keeps flowing, the external load is gradually distributed to the solid skeleton, which starts to deform.

Different theories which model consolidation have been developed in the literature [Biot 1941; Terzaghi 1943; Heinrich and Desoyer 1961], however the two due to Terzaghi and Biot are surely the most widespread ones. Actually, as it was noted also by de Boer [1996], the derivation of the Terzaghi differential equation [Terzaghi 1923] is obscure and essentially driven by the comparison between the phenomenon of soil consolidation and that of heat propagation, rather than from the statement of suitable mechanical principles. On the other hand Biot's theory seems to be well grounded from the mechanical point of view, even if directly restricted to the case of linear elasticity, in its earliest presentation [Biot 1941]. The two models collapse one into the other when considering one-dimensional problems; however, this is not the case when modeling, for instance, the behavior of a saturated porous slab [Mandel 1953] or a saturated porous sphere [Cryer 1963]. In both circumstances Biot's three-dimensional model provides time increasing values of the water pressure (and fluid mass density) at the center of the slab (or sphere) if the Lamé constant $\mu$ of the skeleton is different from zero; on the other hand the solution for $\mu = 0$ coincides with the one derived from Terzaghi's three dimensional model. This localized pore-fluid segregation is known in the literature as the Mandel–Cryer effect.

The occurrence of compaction localization phenomena has recently been discovered by Mollema and Antonellini [1996], who presented evidence of so called "compaction bands" in outcrops of the Jurassic Navajo sandstone in the Kaibab monocline, in Utah. These bands are characterized by volume loss due to microfracturing, but essentially no grain crushing or comminution. Later on, laboratory experiments have been developed [Olsson and Holcomb 2000; 2003] using triaxial compression tests to reproduce the formation of these bands. These tests proved that increasing axial stress $\sigma_{11}$ initially determines only homogeneous axial strain $\epsilon_{11}$, however, when a suitable stress threshold has been overwhelmed, tabular zones associated with nonhomogeneous strain can be detected close to the axial borders of the specimen. Nonuniform compaction also affects fluid flow in the porous material, being detrimental if permeability of the compacted material is much reduced with respect to the uncompacted zone.

The second gradient poromechanical model presented in Part I [Sciarra et al. 2008] is capable of describing fluid mass density boundary segregation even in the one dimensional model. Both in the vicinity of the external consolidating surface and the impermeable wall, suitable boundary layer effects can be predicted by the second gradient model. Formation of segregation bands enhances high gradients of density of the fluid entrapped in the pores of the solid skeleton. This can be explained by means of nonvanishing hyperstresses at the boundary (see [Sciarra et al. 2008, Equations $(32)_2$–$(32)_3$]); these last cause the pore pressure to differ from its reference initial value in a transient period, when dissipation does not yet dominate the evolution process.

Having in mind the classical Terzaghi's consolidation problem, whose space-time evolution is governed by the same equation as that of heat conduction, replacing temperature with pore-pressure, we claim that the second gradient model, presented in [Sciarra et al. 2008], is capable of regularizing the behavior of the Darcy flow inside the porous medium. Because of second gradient effects, the fluid mass density diffusion is smoothed. It is self evident that Terzaghi's theory does not model those phenomena occurring at the boundaries which oppose the fluid flow, for example, pore closure, solid-fluid capillarity, etc. The present model tries to macroscopically account for some of them and aims to establish the preliminary theoretical framework necessary for conceiving and designing any kind of experimental activity. In this paper it is shown that the overpressure occurring at the impermeable boundary actually depends on second gradient coefficients; therefore a more detailed analysis of these effects is recognized to be necessary.

Pore fluid segregation is probably the triggering effect for vertical drain or sand volcano formation, observed after liquefaction [Kolymbas 1998]; these last can indeed be interpreted as bifurcation modes of consolidation, which, in the case of the second gradient model should correspond to the boundary layer detected close to the impermeable wall becoming larger and larger. As it is not possible to identify this bifurcation mode in the case of linearized (small strain) theory, only the limit condition describing the stability/instability limit is detected. Further developments will be devoted to study of the one-dimensional nonlinear problem.

From the mathematical point of view the present model, in which both second gradient conservative (relative to the behavior of the porous solid skeleton) and dissipative (relative to the flow of the saturating fluid) contributions are taken into account, implies that the newly formulated initial boundary value problem of consolidation fits in the framework of the theory of linear pencils of ordinary differential operators on a finite interval with boundary conditions depending on the spectral parameter. We refer to the general results presented, for example, in [Shkalikov 1986; Shkalikov and Tretter 1996; Marletta et al. 2003] for more details on this topic.

## 2. Linearization of the one-dimensional differential problem

We study here the aforementioned consolidation problem referring to the equations of motion for a second gradient porous medium as obtained in [Sciarra et al. 2008, Equations (30) and (31)], restricting our attention to the one-dimensional case. This will allow us to compare our results with the classical ones due to Terzaghi [1923]. Of course, Equations (30)–(31) can also be applied to treat three-dimensional problems, so extending classical Biot's equations.

Clearly, because of the one-dimensional hypothesis, all the gradient and divergence operations appearing in Equations (30) and (31) become simple derivatives with respect to the space variable $x$, and the deformation tensor $\varepsilon$ simply reduces to its only nonzero component $\varepsilon_{xx}$ along the $x$ axis. In the following we will indicate the component $\varepsilon_{xx}$ simply by $\varepsilon$.

From now on we assume the hypothesis of small deformations in the neighborhood of a suitable solid skeleton reference configuration. For the sake of simplicity we will therefore use the same notation as used in [Sciarra et al. 2008] for $\varepsilon$ and $m_f$ to indicate the corresponding incremental quantities with respect to the considered small deformation parameter.

In accordance with the aforementioned assumptions and the introduced nomenclature, the quadratic expression for the Hemholtz free energy density $\Psi$, in terms of the state parameters $(\varepsilon, m_f, \varepsilon^I, m_f^I)$ is adopted as,

$$\Psi = -p_0^{\text{ext.}}\varepsilon + \mu_0^{\text{ext.}} m_f + \frac{1}{2}\big(\lambda + 2\mu + b^2 M\big)\varepsilon^2 + \frac{1}{2} M \Big(\frac{m_f}{m_f^0}\Big)^2$$
$$- bM\varepsilon \frac{m_f}{m_f^0} + \frac{1}{2}\big(K_{ss} + \mathbb{M} K_{sf}^2\big)(\varepsilon^I)^2 + \mathbb{M} K_{sf} \frac{m_f^I}{m_f^0}\varepsilon^I + \frac{1}{2}\mathbb{M}\Big(\frac{m_f^I}{m_f^0}\Big)^2, \quad (1)$$

where $\varepsilon^I$ and $m_f^I$ indicate the first spatial derivatives of $\varepsilon$ and $m_f$, respectively.

The constant coefficients $p_0^{\text{ext.}}$, $\mu_0^{\text{ext.}}$, and $m_f^0$ account for the state of stress of the solid skeleton, the chemical potential of the fluid, and the initial apparent density of the fluid before any external perturbation is applied to the porous system. Moreover, $\lambda$ and $\mu$ are the classical Lamé coefficients, $b$ and $M$ the Biot parameters, and $\mathbb{M}$, $K_{ss}$, and $K_{sf}$ are the second gradient constitutive parameters.

The nonstandard energetic contributions associated with $(\varepsilon^I)^2$, $(m_f^I)^2$, and $\varepsilon^I m_f^I$ are those responsible for the presence of hyperstresses in the balance equations of the overall material and the pure fluid. They allow for describing the compaction/dilatancy localization effects arising in the fluid-filled porous material when the fluid remains entrapped in the solid skeleton. In particular $K_{ss}$ and $\mathbb{M}$ provide nonvanishing hyperstress on the solid skeleton and the pure fluid if second gradient coupling is negligible. Following the interpretation of double forces given in [Sciarra et al. 2008, Section 2], these two constitutive parameters allow for describing internal actions working on the rate of dilatancy along the outward unit normal. The coupling coefficient $K_{sf}$ is labeled as the cocapillarity coefficient in analogy to standard second gradient theories for capillarity models [Seppecher 1987]. It describes second gradient solid-fluid interactions and can be assumed as vanishing in contrast with $K_{ss}$ and $\mathbb{M}$, which will be proved to be positive when positiveness of strain energy density is required (see Equation (2)).

Values of the Lamé and Biot moduli can easily be recovered from the literature [Coussy 2004]; on the other hand no identification for the second gradient moduli is available up to now. It is not the purpose of this paper to set up a constitutive identification based on experiments or mathematical homogenization; conversely, our aim is that of exhibiting the capability of the model presented in [Sciarra et al. 2008] to catch compaction/dilatancy effects. Second gradient parameters therefore will be tuned so as to permit the one-dimensional model to show boundary layer effects for the solid strain and the fluid mass density in the vicinity of the external surfaces.

Requiring definite positiveness of the energy density $\Psi$ defined in Equation (1), the following conditions on the parameters must hold:

$$\lambda + 2\mu > 0, \quad M > 0, \quad \text{and} \quad K_{ss} > 0, \quad \mathbb{M} > 0. \tag{2}$$

The first two conditions are well known in the framework of the classical Biot poromechanics; the second ones restrict the constitutive assumptions on the second gradient parameters.

**2.1.** *Equations of motion.* We will now deduce the linearized form of the equations of motion for the second gradient consolidation problem. In order to do so, it is worthwhile to recall that in the one-dimensional, linearized problem the following chains of equalities hold:

$$\mathbf{F}_s \simeq \mathbf{I} + \nabla_s \mathbf{u} = (1 + \varepsilon)\,\mathbf{I}, \tag{3}$$

$\mathbf{u}$ being the infinitesimal solid displacement field,

$$\mathbf{F}_s^{-1} \simeq (1 - \varepsilon)\,\mathbf{I}, \tag{4}$$

and

$$J_s \simeq \det \mathbf{F}_s = 1 + \text{tr}(\nabla_0 \mathbf{u}) = 1 + \varepsilon, \tag{5}$$

where we recall that all the considered fields in the right hand side of (3)–(5) have to be regarded as incremental quantities with respect to the small deformation parameter. Taking into account (1) for the strain energy $\Psi$ and the one-dimensional form of (3) and (5), the linearized governing equations given in [Sciarra et al. 2008, Equations (30) and (31)], reduce to

$$\left[\left(-p_0^{\text{ext.}} + \lambda + 2\mu + b^2 M\right)\varepsilon - bM\frac{m_f}{m_f^0} - \left(K_{ss} + \mathbb{M}K_{sf}^2\right)\varepsilon^{II} - \mathbb{M}K_{sf}\frac{m_f^{II}}{m_f^0}\right]^I = 0, \tag{6}$$

and

$$-m_f^0\left[M\frac{m_f^I}{\left(m_f^0\right)^2} - \frac{bM}{m_f^0}\varepsilon^I - \frac{\mathbb{M}K_{sf}}{m_f^0}\varepsilon^{III} - \frac{\mathbb{M}}{\left(m_f^0\right)^2}m_f^{III}\right] - D(v_f - v_s) + \left[\alpha(v_f - v_s)^I\right]^I = 0, \tag{7}$$

for the solid skeleton and the pure fluid, respectively. We have denoted by $D$ and $\alpha$ the only nonzero component of the Darcy and Darcy-like tensors $\mathbb{D}$ and $\mathbb{A}$, respectively (see [Sciarra et al. 2008, Equation (25)]), and by $v_f - v_s$ the vertical component of the relative velocity. In the absence of inertia forces the solid momentum conservation law, Equation (6), can be integrated in the form

$$\left(\lambda + 2\mu + b^2 M - p_0^{\text{ext.}}\right)\varepsilon - bM\frac{m_f}{m_f^0} - \left(K_{ss} + \mathbb{M}K_{sf}^2\right)\varepsilon^{II} - \mathbb{M}K_{sf}\frac{m_f^{II}}{m_f^0} = \text{const.} := c_0. \tag{8}$$

Moreover, considering that $v_f - v_s$ is related to the apparent fluid density $m_f$ by means of the linearized continuity equation $\dot{m}_f + m_f^0(v_f - v_s)^I = 0$, (7) can be rewritten, performing a derivative with respect to the space variable $x$, as

$$-m_f^0\left[M\frac{m_f^{II}}{\left(m_f^0\right)^2} - \frac{bM}{m_f^0}\varepsilon^{II} - \frac{\mathbb{M}K_{sf}}{m_f^0}\varepsilon^{IV} - \frac{\mathbb{M}}{\left(m_f^0\right)^2}m_f^{IV}\right] + \frac{D}{m_f^0}\dot{m}_f - \frac{\alpha}{m_f^0}\dot{m}_f^{II} = 0, \quad (9)$$

where we have indicated by $\dot{m}_f$ the time derivative of $m_f$.

In order to write the linearized governing equations in a dimensionless form, the following quantities are introduced:

$$\xi = \frac{x}{L}, \qquad \tilde{m}_f = \frac{m_f}{m_f^0}, \qquad \tilde{t} = \frac{t}{\tau}, \quad \text{with} \quad \tau = \frac{DL^2}{M},$$

where $L$ is the depth of the solid layer.

According to these definitions, (8) and (9) can be rewritten in their dimensionless form as

$$\frac{(\lambda + 2\mu + b^2M - p_0^{\text{ext.}})}{\lambda + 2\mu}\varepsilon - \frac{bM}{\lambda + 2\mu}\tilde{m}_f - \frac{(K_{ss} + \mathbb{M}K_{sf}^2)}{(\lambda + 2\mu)L^2}\varepsilon^{II} - \frac{\mathbb{M}K_{sf}}{(\lambda + 2\mu)L^2}\tilde{m}_f^{II} = \frac{c_0}{\lambda + 2\mu}, \quad (10)$$

and

$$\frac{\mathbb{M}}{ML^2}\tilde{m}_f^{IV} + \frac{\mathbb{M}}{ML^2}K_{sf}\varepsilon^{IV} - \tilde{m}_f^{II} + b\varepsilon^{II} - \frac{\alpha}{DL^2}\dot{\tilde{m}}_f^{II} + \dot{\tilde{m}}_f = 0, \quad (11)$$

which represent the linearized equations of motion for the consolidation problem. For the sake of simplicity, we will no longer distinguish between $m_f$ and $\tilde{m}_f$, and, if not specified, $m_f$ will indicate the dimensionless quantity. Moreover, the dimensionless variables $\xi$ and $\tilde{t}$ will be also indicated by $x$ and $t$ if no confusion can arise.

**2.2. Boundary conditions.** The constant $c_0$ is deduced from the boundary condition (BC) in $x = 0$ given in [Sciarra et al. 2008, Equation (32)$_1$] which, in the linearized form, reads

$$\frac{(\lambda + 2\mu + b^2M - p_0^{\text{ext.}})}{\lambda + 2\mu}\varepsilon - \frac{bM}{\lambda + 2\mu}m_f - \frac{\left(K_{ss} + \mathbb{M}K_{sf}^2\right)}{(\lambda + 2\mu)L^2}\varepsilon^{II} - \frac{\mathbb{M}K_{sf}}{(\lambda + 2\mu)L^2}m_f^{II} = -\frac{\Delta p^{\text{ext.}}}{(\lambda + 2\mu)};$$

here $\Delta p^{\text{ext.}}$ represents the incremental external pressure acting on the system, deriving from the linearization process ($p^{\text{ext.}} = p_0^{\text{ext.}} + \Delta p^{\text{ext.}}$). In other words, $\Delta p^{\text{ext.}}$ is the perturbation in the external load applied on the surface of the soil layer. Comparing this BC with (10) it is easy to recognize that $c_0 = -\Delta p^{\text{ext.}}$.

Equations (10) and (11) represent a differential system of the sixth order in the space variable $x$ and of the first order in time, the integration of which requires therefore six boundary conditions and one initial condition. In classical poromechanics the Terzaghi consolidation problem does not take into account second gradient effects, and indeed it can be obtained from (10) and (11) when the second gradient parameters $\mathbb{M}$, $K_{ss}$, $K_{sf}$, and $\alpha$ are vanishing. Clearly, the problem reduces in this case to a second order system with respect to the space variable $x$.

The boundary conditions for the consolidation problem are derived from the general ones deduced in [Sciarra et al. 2008, Equation (32)]. In particular, since the given problem is one-dimensional, surface divergence and surface gradient operations do not contribute to the BCs; moreover no edge $\mathfrak{E}_k$ of the boundary exists. Extending the classical BCs stated in the Terzaghi consolidation problem we assume:

- Zero fluid traction in $x = 0$. This BC states that the surface of the solid layer is kept drained, meaning that the fluid reaching the surface is continuously removed from the surface itself. This BC corresponds to the one given in [Sciarra et al. 2008, Equation $(32)_1$] which, in its linearized, dimensionless form, reduces to

$$m_f - b\varepsilon - \frac{\mathbb{M}K_{sf}}{ML^2}\varepsilon^{II} - \frac{\mathbb{M}}{ML^2}m_f^{II} + \frac{\alpha}{DL^2}\dot{m}_f = \frac{m_f^0 \Delta\mu^{\text{ext.}}}{M} = 0, \quad (12)$$

where $\Delta\mu^{\text{ext.}}$ represents the incremental chemical potential. In other words we have assumed the linearization $\mu^{\text{ext.}} = \mu_0^{\text{ext.}} + \Delta\mu^{\text{ext.}}$. Assuming that the fluid is continuously removed from the surface of the layer, this implies a restriction to the case $\Delta\mu^{\text{ext.}} = 0$.

- Impermeable soil in $x = L$. With this BC we assume that the relative velocity is equal to zero in $x = L$, implying $v_f - v_s = 0$. Using Equation (7), which holds everywhere in the interval $[0, L]$, the impermeability of the layer $x = L$ can be rewritten in its dimensionless form as

$$-m_f^I + b\varepsilon^I + \frac{\mathbb{M}K_{sf}}{ML^2}\varepsilon^{III} + \frac{\mathbb{M}}{ML^2}m_f^{III} - \frac{\alpha}{DL^2}\dot{m}_f^I = 0. \quad (13)$$

- Zero double force for the overall system in $x = 0$ and $x = L$. These BCs are those ones given in [Sciarra et al. 2008, Equation $(32)_2$], and can be rewritten in their linearized dimensionless form as

$$m_f^I + \frac{(K_{ss} + \mathbb{M}K_{sf}^2)}{\mathbb{M}K_{sf}}\varepsilon^I = 0.$$

We remind that the overall double forces are the contact forces introduced in the second gradient model, which work on the rate of pore opening/pore shrinkage. With the assumption of vanishing double forces on the boundary of the porous material, we claim that no external source of double force exists; internal double forces, on the contrary, allow for capturing the effects of pressure gradient concentration in the neighborhood of the external and impermeable surfaces [Holcomb and Olsson 2003].

- Zero fluid double force in $x = 0$ and in $x = L$. These BCs are those ones given in [Sciarra et al. 2008, Equation $(32)_3$]. They can be rewritten in their linearized dimensionless form as

$$m_f^I + K_{sf}\varepsilon^I = 0. \quad (14)$$

The assumption on fluid double forces can be interpreted similarly to that considered for the overall double forces. In this case no external double forces are exerted on the fluid boundary, but internal double forces, associated with pressure gradient concentration, account for internal capillarity and wetting phenomena.

## 3. Initial boundary value problem

The differential Equations (10) and (11) can be reduced to a unique differential equation introducing an auxiliary function $V(x,t)$ which satisfies the relationships

$$\varepsilon = \frac{K_{sf}\mathbb{M}}{(\lambda + 2\mu)L^2}V^{II} + \frac{bM}{(\lambda + 2\mu)}V, \tag{15}$$

and

$$m_f = \frac{(\lambda + 2\mu + b^2 M - p_0^{\text{ext.}})}{(\lambda + 2\mu)}V - \frac{K_{ss} + \mathbb{M}K_{sf}^2}{(\lambda + 2\mu)L^2}V^{II} + \frac{\Delta p^{\text{ext.}}}{bM}. \tag{16}$$

In such a way, Equation (10) is identically satisfied, while (11) can be rewritten, after some straightforward calculations, as

$$C_1 V^{VI} - C_2\left(p_0^{\text{ext.}}\right)V^{IV} - C_3 \dot{V}^{IV} + C_4\left(p_0^{\text{ext.}}\right)V^{II} + C_5\left(p_0^{\text{ext.}}\right)\dot{V}^{II} - C_6\left(p_0^{\text{ext.}}\right)\dot{V} = 0; \tag{17}$$

on the other hand, the boundary conditions given in (12)–(14) read

$$\mathbb{C}_1 V^{IV} - \mathbb{C}_2\left(p_0^{\text{ext.}}\right)V^{II} - \mathbb{C}_3 \dot{V}^{II} + \mathbb{C}_4\left(p_0^{\text{ext.}}\right)V + \mathbb{C}_5\left(p_0^{\text{ext.}}\right)\dot{V} + \mathbb{C}_6 = 0 \quad \text{at } x = 0, \tag{18}$$

$$V^I = 0 \quad \text{at } x = 0, L, \qquad V^{III} = 0 \quad \text{at } x = 0, L, \qquad V^V = 0 \quad \text{at } x = L. \tag{19}$$

Finally, the initial condition (corresponding to the instant in which the external load is applied) is deduced assuming that the apparent Lagrangian fluid density is vanishing for $t = 0^+$. For instance, $m_f(x, t = 0^+) = 0$; this initial condition states, similarly to in classical consolidation, that, at the instant in which the external load is applied there is no instantaneous variation of the fluid density $m_f$ inside the soil layer. In terms of the auxiliary function $V$ the initial datum reads as

$$V(x, 0^+) := V_{in} = -\frac{\Delta p^{\text{ext.}}}{bM}\frac{1}{C_4(\pi_0) + k_6}, \tag{20}$$

where (16) with $m_f = 0$ has been solved using BCs given in (19).

All the coefficients appearing in the governing equation, (17), as well as in the initial and boundary conditions, (20) and (19), depend on the constitutive parameters, the solid initial stress $p_0^{\text{ext.}}$, and the increment of the external force $\Delta p^{\text{ext.}}$; their expressions are listed in Appendix A.

It must be remarked that the expression for the energy density $\Psi$ assumed in (1) would not allow the linearized differential problem to explicitly depend on the initial solid stress $p_0^{\text{ext.}}$ and on the initial chemical potential $\mu_0^{\text{ext.}}$. In fact, considering (1) we can write, in dimensionless form,

$$\frac{\partial \Psi}{\partial \varepsilon} = \frac{1}{\lambda + 2\mu}p_0^{\text{ext.}} + (1 + k_6)\varepsilon - bk_5 m_f, \qquad \frac{\partial \Psi}{\partial m_f} = \frac{m_f^0}{\lambda + 2\mu}\mu_0^{\text{ext.}} + k_5 m_f - bk_5 \varepsilon, \tag{21}$$

so that there are no linear terms (in $\varepsilon$ and $m_f$) involving $p_0^{\text{ext.}}$ and $\mu_0^{\text{ext.}}$ coming from $\partial \Psi/\partial \varepsilon$ and $\partial \Psi/\partial m_f$. On the other hand the dependence of the differential system (17)–(19) on $p_0^{\text{ext.}}$ is due to so called geometrical nonlinearities; as matter of fact it is the presence of $\mathbf{F}_s$ in the balance of the total momentum (see [Sciarra et al. 2008]) which even in linearized problems implies a nontrivial dependence of the governing equations on $p_0^{\text{ext.}}$ (see the term $\varepsilon p_0^{\text{ext.}}$ in (10)).

Considering the linearity of the differential problem and the nonhomogeneity appearing in the BC, given in Equation (18), we will look for a solution $V(x, t)$ in the form

$$V(x, t) = \bar{V}(x) + W(x, t), \tag{22}$$

where $\bar{V}(x)$ is the solution of the stationary problem

$$C_1 \bar{V}^{VI} - C_2 \bar{V}^{IV} + C_4 \bar{V}^{II} = 0, \tag{23}$$

with nonhomogeneous BCs

$$\mathbb{C}_1 \bar{V}^{IV} - \mathbb{C}_2 \bar{V}^{II} + \mathbb{C}_4 \bar{V} = -\mathbb{C}_6 \quad \text{at } x = 0, \tag{24}$$

$$\bar{V}^I = 0 \quad \text{at } x = 0, L, \qquad \bar{V}^{III} = 0 \quad \text{at } x = 0, L, \qquad \bar{V}^V = 0 \quad \text{at } x = L, \tag{25}$$

while the deviation $W(x, t)$ is the solution of the initial boundary value problem (IBVP)

$$C_1 W^{VI} - C_2 W^{IV} - C_3 \dot{W}^{IV} + C_4 W^{II} + C_5 \dot{W}^{II} - C_6 \dot{W} = 0, \tag{26}$$

with homogeneous BCs

$$\mathbb{C}_1 W^{IV} - \mathbb{C}_2 W^{II} - \mathbb{C}_3 \dot{W}^{II} + \mathbb{C}_4 W + \mathbb{C}_5 \dot{W} = 0 \quad \text{at } x = 0, \tag{27}$$

$$W^I = 0 \quad \text{at } x = 0, L, \qquad W^{III} = 0 \quad \text{at } x = 0, L, \qquad W^V = 0 \quad \text{at } x = L, \tag{28}$$

and nontrivial initial condition

$$W(x, 0^+) := W_{in} = V_{in} - \bar{V}(x). \tag{29}$$

For the sake of simplicity we will no longer specify the dependence of the coefficients $C_i$ and $\mathbb{C}_i$ on the prestress parameter $p_0^{\text{ext.}}$.

It is easy to prove that the solution of the stationary problem given by (23)–(25) when $\mathbb{C}_4 \neq 0$ is given by

$$\bar{V}(x) = -\frac{\mathbb{C}_6}{\mathbb{C}_4}, \qquad \mathbb{C}_4 \neq 0, \tag{30}$$

while if $\mathbb{C}_4 = 0$ the stationary solution $\bar{V}(x)$ exists if and only if $\mathbb{C}_6 = 0$; in this case a family of constant solutions of the stationary problem arises, so that we can write $\bar{V}(x) = K$, and $\mathbb{C}_4 = \mathbb{C}_6 = 0$, where K is an undetermined constant. On the basis of the preliminary study of the stationary solution $\bar{V}(x)$ we can state that, according to the assumption (22), a solution of the given problem for the variable $V$ exists if and only if $\mathbb{C}_4 \neq 0$ or $\mathbb{C}_4 = \mathbb{C}_6 = 0$. We will now restrict our attention to the case $\mathbb{C}_4 \neq 0$ and will analyze the case $\mathbb{C}_4 = 0$ later.

## 4. Fourier series solution

The initial boundary value problem given by (26)–(29) is solved using the method of separation of variables; in other words we assume that $W(x, t) = X(x)T(t)$. A straightforward calculation yields to the definition of the real eigenparameter $\lambda$ as $\lambda = \dot{T}/T$, which leads to $T(t) = T_0 e^{\lambda t}$. Consequently, the function $X(x)$ must satisfy the eigenvalue problem

$$C_1 X^{VI} - C_2 X^{IV} + C_4 X^{II} = \lambda \left( C_3 X^{IV} - C_5 X^{II} + C_6 X \right), \tag{31}$$

endowed with the BCs

$$\mathbb{C}_1 X^{IV} - \mathbb{C}_2 X^{II} + \mathbb{C}_4 X = \lambda \left( \mathbb{C}_3 X^{II} - \mathbb{C}_5 X \right), \quad \text{at } x = 0, \tag{32}$$

$$X^I = 0 \quad \text{at } x = 0, L, \qquad X^{III} = 0 \quad \text{at } x = 0, L, \qquad X^V = 0 \quad \text{at } x = L. \tag{33}$$

This is a nonclassical spectral problem since the BCs also depend on the spectral parameter $\lambda$; in the literature this kind of spectral problem is referred to as a linear pencil $\mathbb{L}(X) = \lambda \mathbb{A}(X)$. Many authors investigate the spectral properties of the differential operators $\mathbb{L}$ and $\mathbb{A}$ in suitable function spaces in order to guarantee completeness and orthonormality for the eigenfunction system and discreteness of the spectrum [Shkalikov 1986; Shkalikov and Tretter 1996; Marletta et al. 2003]. Here we rely on these general results and numerically determine a subset of the eigenfunction space so as to approach the requirements of the Parseval equality [Kolmogorov and Fomin 1975].

According to the aforementioned properties of the eigensystem, the solution of the considered IBVP can be given in Fourier series form as

$$W(x, t) = \sum_{k=0}^{+\infty} p_k X_k(x) e^{\lambda_k t}, \tag{34}$$

where $p_k$ denotes the $k$-th Fourier coefficient, and, in particular, $p_0$ the Fourier coefficient relative to the null eigenvalue $\lambda = 0$ (if any). It is easy to prove that if $\mathbb{C}_4 \neq 0$ the eigenfunction $X_0$ relative to the null eigenvalue is the trivial one $X_0 = 0$, so that in Equation (34) $k$ runs now from one to infinity.

The eigenfunctions $(X_k)_{k \in \mathbb{N}}$ are orthogonal with respect to the following bilinear form defined, in the Hilbert space $H^3([0, L]) \times H^3([0, L])$, as

$$\langle X_k, X_h \rangle := \alpha_0 \int_0^L X_k X_h dx + \alpha_1 \int_0^L X_k^I X_h^I dx + \alpha_2 \int_0^L X_k^{II} X_h^{II} dx + \alpha_3 \int_0^L X_k^{III} X_h^{III} dx, \tag{35}$$

where the coefficients $\alpha_i$ are defined as

$$\alpha_0 = \mathbb{C}_4 C_6, \qquad\qquad \alpha_1 = \mathbb{C}_4 C_5 - C_4 \mathbb{C}_5 + \mathbb{C}_2 C_6,$$

$$\alpha_2 = \mathbb{C}_4 C_3 - \mathbb{C}_3 C_4 + \mathbb{C}_2 C_5 - C_2 \mathbb{C}_5 + \mathbb{C}_1 C_6, \qquad \alpha_3 = \mathbb{C}_2 C_3 - C_2 \mathbb{C}_3 + \mathbb{C}_1 C_5 - C_1 \mathbb{C}_5.$$

It must be noted that expression (35) is indeed an inner product over the aforementioned function space if and only if all the coefficients $\alpha_i$ are positive definite.

We notice that when the initial stress $p_0^{\text{ext.}}$ is vanishing the $\alpha_i$ coefficients are all positive (assuming positiveness of the energy $\Psi$, see (2)), so that (35) always represents an inner product. On the other hand, it is easy to verify that in the presence of prestress the positiveness of the aforementioned coefficients $\alpha_i$ is guaranteed if and only if $p_0^{\text{ext.}} < \lambda + 2\mu$ ($\Leftrightarrow \mathbb{C}_4 > 0$).

The explicit form of the Fourier coefficients $p_k$ is determined by projecting the initial datum (29) on the $k$-th element of the Fourier series according to the inner product (35); in particular we can find

$$\langle W_{in}, X_k \rangle = \alpha_0 W_{in} \int_0^L X_k dx, \tag{36}$$

$$\langle W_{in}, X_k \rangle = \left\langle \sum_{h=1}^{+\infty} p_h X_h, X_k \right\rangle = \sum_{h=1}^{+\infty} p_h \langle X_h, X_k \rangle = p_k \|X_k\|^2, \tag{37}$$

where we have noted by $\|\cdot\| = \langle \cdot, \cdot \rangle^{1/2}$ the norm induced by the inner product $\langle \cdot, \cdot \rangle$. Comparing Equation (36) with (37) it is easy to recognize that

$$p_k = \frac{\alpha_0 W_{in} \int_0^L X_k \, dx}{\|X_k\|^2},$$

so that, recalling (34), the final form of the solution is

$$W(x,t) = \alpha_0 W_{in} \sum_{k=1}^{+\infty} \left[ \frac{1}{\|X_k\|^2} \int_0^L X_k(\xi) d\xi \right] X_k(x) e^{\lambda_k t},$$

and, according to (22) and (30), the solution for the variable $V(x,t)$ is finally given by

$$V(x,t) = -\frac{\mathbb{C}_6}{\mathbb{C}_4} + \alpha_0 W_{in} \sum_{k=1}^{+\infty} \left[ \frac{1}{\|X_k\|^2} \int_0^L X_k(\xi) d\xi \right] X_k(x) e^{\lambda_k t}. \quad (38)$$

Finally, the fields $\varepsilon$ and $m_f$ can be evaluated using (38) with (15) and (16) respectively.

**4.1.** *The limit case $\mathbb{C}_4 = 0$.* We have already mentioned that when $\mathbb{C}_4 = 0$ ($p_0^{\text{ext.}} = \lambda + 2\mu$) the stationary solution $\bar{V}(x)$ exists if and only if $\mathbb{C}_6 = 0$ ($\Leftrightarrow \Delta p^{\text{ext.}} = 0$), and it is an undetermined constant $K$. This means that, corresponding to a critical value of the prestress $p_0^{\text{ext.}}$, no solution can be found if perturbing the porous system with an external load $\Delta p^{\text{ext.}}$. The only possible solution is relative to the unloaded configuration of the porous system ($\Delta p^{\text{ext.}} = 0$). In this case the solution for $V(x,t)$ is found by solving the differential problem given by (17), (19), and (20) when $\mathbb{C}_4 = \mathbb{C}_6 = 0$. Separating the variables, $V(x,t) = X(x)T(t)$, the solution can be found in Fourier series form as

$$V(x,t) = \sum_{k=0}^{+\infty} p_k X_k(x) e^{\lambda_k t}. \quad (39)$$

It must be noticed that when $\mathbb{C}_4 = 0$ the inner product (35) reduces to

$$\langle X_k, X_h \rangle_{\mathbb{C}_4=0} := \alpha_1 \int_0^L X_k^I X_h^I \, dx + \alpha_2 \int_0^L X_k^{II} X_h^{II} \, dx + \alpha_3 \int_0^L X_k^{III} X_h^{III} \, dx,$$

and it is still well defined over the quotient space of the $H^3([0, L])$ functions, differing at most by a constant. It follows that the Fourier coefficients $p_k$ are now determined on the basis of the reduced form $\langle \cdot, \cdot \rangle_{\mathbb{C}_4=0}$ of the inner product, according to the identities involving the initial condition $V_{in} = $ constant,

$$0 = \langle V_{in}, X_0 \rangle_{\mathbb{C}_4=0} = \left\langle p_0 X_0 + \sum_{k=1}^{+\infty} p_k X_k(x), X_0 \right\rangle_{\mathbb{C}_4=0} = p_0 \langle X_0, X_0 \rangle_{\mathbb{C}_4=0}, \quad (40)$$

$$0 = \langle V_{in}, X_k \rangle_{\mathbb{C}_4=0} = \left\langle p_0 X_0 + \sum_{k=1}^{+\infty} p_h X_h(x), X_k \right\rangle_{\mathbb{C}_4=0} = p_k \|X_k\|_{\mathbb{C}_4=0}^2, \quad \text{for all } k \in \mathbb{N}, \quad (41)$$

where we have noted by $\|\cdot\|_{\mathbb{C}_4=0} = \langle \cdot, \cdot \rangle_{\mathbb{C}_4=0}^{1/2}$ the norm induced by the inner product $\langle \cdot, \cdot \rangle_{\mathbb{C}_4=0}$. Notice that $p_0$ and $X_0 = \bar{V} = K$ are the Fourier coefficient and the constant eigenfunction corresponding to the null eigenvalue $\lambda_0 = 0$, respectively, while $(X_k)_{k \in \mathbb{N}}$ are the remaining eigenfunctions. Since $X_0 =$ constant, Equation (40) reads $p_0 0 = 0 \implies p_0$ undetermined, moreover, (41) gives $p_k \|X_k\|^2_{\mathbb{C}_4 = 0} = 0 \implies p_k = 0, \forall k \in \mathbb{N}$. According to (39), the solution for $V(x, t)$ is an undetermined constant, so $V(x, t) = p_0 X_0 := p_0 K = $ constant.

We want to remark that all the Fourier coefficients $p_k$ corresponding to nonvanishing eigenvalues turn to be zero only because the initial condition $V_{in}$ has been assumed to be constant; if it was not the case, (41) would have stated the expression for the coefficients $p_k$, and the solution for $V(x, t)$ would have been known except for a constant $K$. The fact that a family of constant solutions for $V(x, t)$ arises can be seen as a sort of bifurcation phenomenon which is triggered when $p_0^{\text{ext.}}$ reaches the critical value $\lambda + 2\mu$.

Finally, we underline that the null eigenvalue $\lambda_0 = 0$ belongs to the spectrum of the differential problem only when $\mathbb{C}_4 = 0$; in the following section we will show that when $\mathbb{C}_4 > 0$ only negative eigenvalues exist, while if $\mathbb{C}_4 < 0$ some positive eigenvalues appear.

## 5. Numerical results

In this section we will show the numerical solution of the differential problem, given by (17)–(20), for a particular set of values of the constitutive parameters, which are listed in Table 1. The first gradient parameters are those relative to a water saturated clay, while the values of the second gradient dimensionless numbers are chosen in order to let boundary layer effects arise.

Fixing suitable values for the initial external pressure ($p_0^{\text{ext.}} = 4.9$ GPa) and for the increment of this latter ($\Delta p^{\text{ext.}} = 1$ MPa), so as to guarantee $\mathbb{C}_4 > 0$ and $\mathbb{C}_6 \neq 0$, we look for a numerical solution $V(x, t)$ given by (38). In particular, we look for a numerical solution $X(x)$ of the differential problem, given by (31)–(33), in the form

$$X(x) = \sum_{i=1}^{6} K_i e^{\beta_i(\lambda) x}, \tag{42}$$

where $K_i$ are the integration constants and $\beta_i(\lambda)$ are the solutions of the characteristic polynomial associated with the differential equation, (31). Consequently, BCs given by (32)–(33) yield

$$\mathbf{A}(\lambda)\mathbf{v} = 0, \tag{43}$$

where $\mathbf{A}(\lambda)$ is a suitably defined $6 \times 6$ matrix and $\mathbf{v} := (K_1, ..., K_6)$. We notice that the matrix $\mathbf{A}(\lambda)$ depends on the eigenparameter $\lambda$ both because it appears in the differential equation, (31), and in the boundary condition, (32). It follows that the resulting eigenvalue problem cannot be classified as a standard eigenvalue problem. The system of algebraic equations, (43), has a nontrivial solution if and only if $\det [\mathbf{A}(\lambda)] = 0$, which leads to the calculation of the eigenvalues $\lambda_k$ (discrete spectrum). For each eigenvalue $\lambda_k$ an eigenfunction $X_k(x)$ is deduced in the form of Equation (42) so that a numerical solution for $V(x, t)$ can be found according to (38).

| $M(GPa)$ | $\lambda(GPa)$ | $\mu(GPa)$ | $k_1$ | $k_2$ | $k_3$ | $k_4$ |
|---|---|---|---|---|---|---|
| 5 | 2.3 | 1.5 | $10^{-2}$ | $10^{-2}$ | $10^{-2}$ | $10^{-2}$ |

**Table 1.** The values of first gradient elasticity parameters relative to a normally consolidated water saturated clay, together with trial values of second gradient dimensionless parameters.

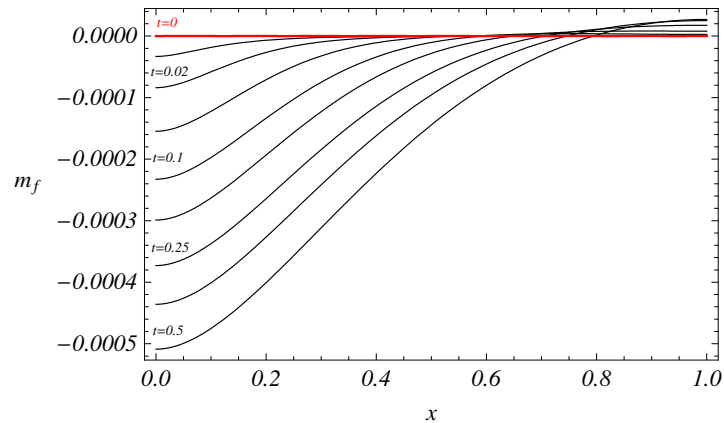

**Figure 1.** Behavior of the fluid mass density $m_f$ versus depth $x$ in the vicinity of the time $t = 0^+$. A segregation of the fluid arises close to the impermeable wall at $x = 1$.

The numerical solution for $V(x, t)$ involves a finite number of eigenfunctions $N$, where the choice of $N$ is made so as to approach the condition stated by the Parseval equality [Kolmogorov and Fomin 1975]. Once the numerical solution for $V(x, t)$ has been found, we can deduce the corresponding solutions for the fields $\varepsilon$ and $m_f$ simply be referring to (15) and (16).

We now show the behavior of the fields $\varepsilon$ and $m_f$ corresponding to the aforementioned values of the constitutive parameters, and initial and incremental pressures. In Figure 1 the fluid apparent density $m_f$ versus $x$ is depicted for times in the very close neighborhood of $t = 0^+$. It can be noticed that a critical depth $x_{cr} \simeq 0.8$ exists such that the density $m_f$ decreases for $0 \leq x < x_{cr}$, while it increases for $x_{cr} < x \leq 1$. This means that the fluid of the upper regions actually leaves the layer, while the fluid contained in the deeper regions remains entrapped in the pores whose deformation consequently increases the apparent density $m_f$. When increasing time (see Figure 2), the apparent density $m_f$ decreases along the whole depth of the layer, and finally approaches a constant value. This means that the fluid starts flowing out also from the deeper regions until the system reaches a new equilibrium and no fluid leaves the layer anymore. This effect is evidently related to viscosity, which dominates the evolution of the fluid density as time becomes larger and larger.

As far as the vertical deformation $\varepsilon$ is concerned, the same qualitative behavior as that of $m_f$ is detected (see Figure 3 and Figure 4). For times close to $t = 0^+$ (see Figure 3) the upper regions of the layer undergo to a vertical compression, which is connected to the fact that less fluid is present in the pores, while the deeper regions experience a sort of dilatancy which is connected to an over pressurization of the saturating fluid.

For increasing times (see Figure 4) a general further compression is detected along the whole depth of the layer (this is due to the fact that the fluid is uniformly flowing along the layer) until the layer does not deform anymore (equilibrium).

We remark that the chosen values of the prestress $p_0^{\text{ext.}}$ are such that $\mathbb{C}_4 > 0$ so that the inner product Equation (35) is well defined. Consequently the solution for $V(x,t)$ (and thus for $\varepsilon$ and $m_f$) can be numerically evaluated.

It is interesting to notice that when the initial stress $p_0^{\text{ext.}}$ is such that $p_0^{\text{ext.}} < \lambda + 2\mu$ ($\mathbb{C}_4 > 0$) only negative eigenvalues $\lambda_k < 0$ have been found, while in the region where $\mathbb{C}_4 < 0$ a finite number of positive eigenvalues arise. In Figure 5 the behavior of the first eigenvalue $\lambda_1$ is shown when varying $p_0^{\text{ext.}}$ through the threshold $p_0^{\text{ext.}} = \lambda + 2\mu$. It is worth noticing that when $\lambda_1$ passes from negative to positive values, the solution $V(x,t)$ given in the form of Equation (38) blows up due to the presence of positive time exponentials; the solution thus experiences an unstable behavior related to the fact that $p_0^{\text{ext.}}$ reaches a critical value. This kind of instability is known as geometrical instability since the presence of $p_0^{\text{ext.}}$ in the differential problem is due to the geometry of the problem (see Equation (21)).

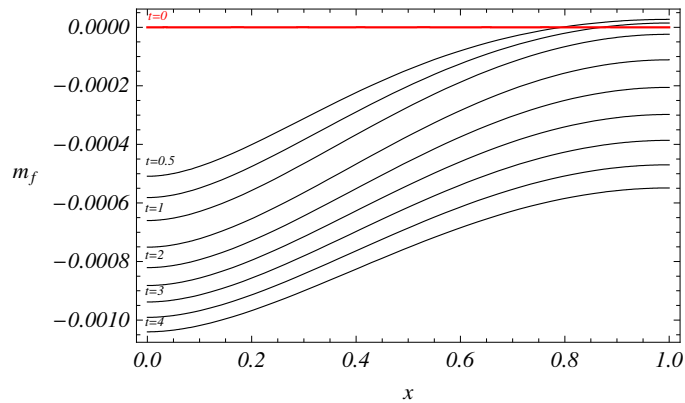

**Figure 2.** Profile of $m_f$ for further times. Notice that $m_f$ tends to assume a constant value for $t \to +\infty$, approaching equilibrium.

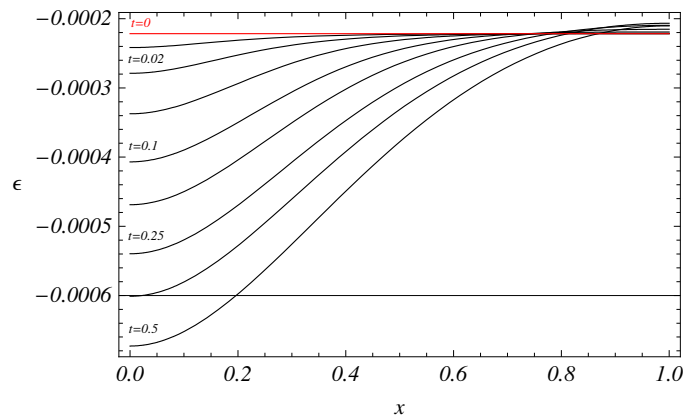

**Figure 3.** Profile of the vertical solid strain $\varepsilon$ versus $x$ for times close to $t = 0^+$. A dilatancy of the solid skeleton is detected in the neighborhood of $x = 1$.

The linearity of the present model does not allow us to capture solutions associated with unstable conditions; this is evident when considering that the bilinear form given in (35) is no longer a well defined inner product.

In order to show the influence of the solid prestress on the behavior of $\varepsilon$ and $m_f$ we have found solutions for different values of $p_0^{\text{ext.}}$ and noticed changes in the solution when approaching the threshold $\mathbb{C}_4 = 0$. Figure 6 shows the behavior of $m_f$ when $p_0^{\text{ext.}}$ progressively approaches the critical value $p_0^{\text{ext.}} = \lambda + 2\mu$. When increasing the value of $p_0^{\text{ext.}}$ the fluid density decreases in the superficial regions of the layer, while increasing in the deeper ones. This means that the initial stress increases the capability of the fluid to flow out from the skeleton matrix close to the external surface, while pumping it in the deeper layers.

Let us now consider the second gradient constitutive parameters and the initial stresses to be vanishing. The resulting differential problem reduces to the classical Terzaghi consolidation problem. More particularly, Equation (10) reduces to

$$\varepsilon = \frac{bM}{\lambda + 2\mu + b^2 M} m_f - \frac{\Delta p^{\text{ext.}}}{\lambda + 2\mu + b^2 M},$$

which, substituted in (11), gives

$$\dot{m}_f = a m_f^{II}, \qquad a = \frac{(\lambda + 2\mu)}{\lambda + 2\mu + b^2 M}. \tag{44}$$

The Terzaghi consolidation problem thus reduces to the differential equation, Equation (44), together with the initial datum $m_f(x, 0^+) = 0$ and the BCs, (12) and (13), which simplify into

$$m_f = \frac{\lambda + 2\mu + b^2 M}{\lambda + 2\mu} \left( b \frac{\Delta p^{\text{ext.}}}{\lambda + 2\mu} \right) := c \qquad \text{at } x = 0. \tag{45}$$

and $m_f^I = 0$ at $x = L$, respectively.

It is easy to notice that the BC, (45), and the initial datum, $m_f(x, 0^+) = 0$, are not consistent, so the Terzaghi solution for $m_f$ exhibits the well known behavior of the classical unidimensional heat equation.

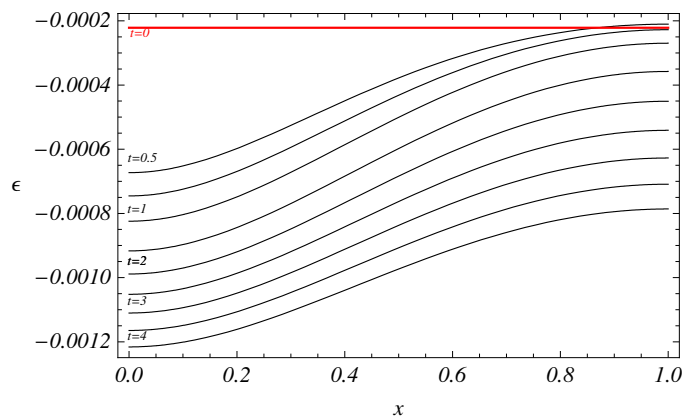

**Figure 4.** Behavior of $\varepsilon$ for further times. Notice that the system tends to reach a state of equilibrium for $t \to +\infty$.

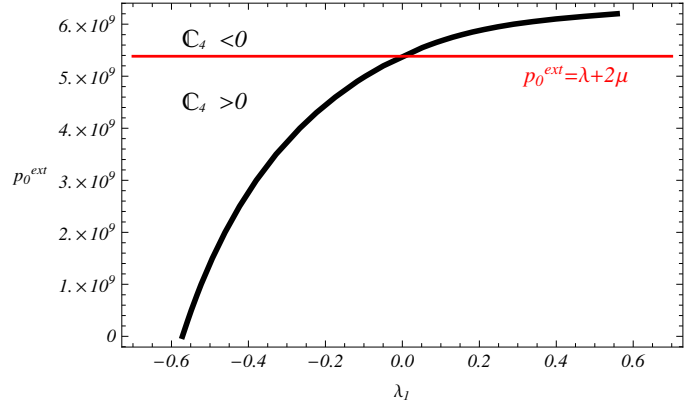

**Figure 5.** The value of the first eigenvalue $\lambda_1$ versus the prestress $p_0^{\text{ext.}}$.

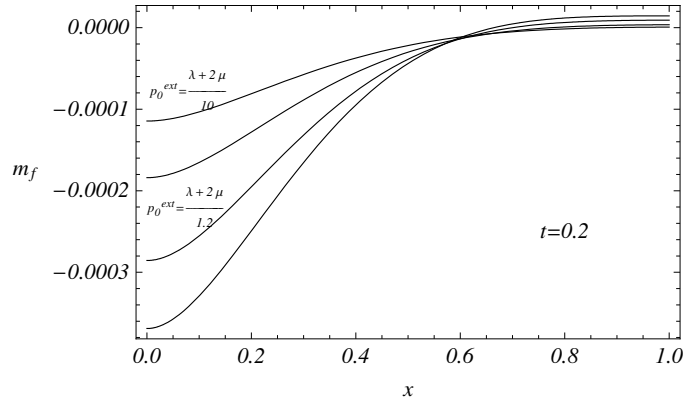

**Figure 6.** $m_f$ profile versus depth for $t = 0.2$ and for different values of the prestress $p_0^{\text{ext.}}$ in the neighborhhod of $p_0^{\text{ext.}} = \lambda + 2\mu$.

As it will be shown in the following, this discontinuity between the initial datum and the BC is cured by the second gradient model. The Terzaghi solution given in terms of the fluid mass density $m_f$ reads

$$m_f(x,t) = c + \sum_{k=1}^{+\infty} \sqrt{2a} \sin\left[\left(\frac{\pi}{2} + k\pi\right)x\right] e^{\lambda_k t}, \qquad \lambda_k = a\left(\frac{\pi}{2} + k\pi\right)^2.$$

In Figure 7 we show the comparison between the Terzaghi and the second gradient solutions for $m_f$ and $\varepsilon$, respectively (in absence of prestresses), corresponding to the initial condition. The Terzaghi solution (blue line) tends to a step function due to the discontinuity between the initial datum and the BC; on the other hand this discontinuity is not present in the second gradient solution (red line). Moreover, we underline that the second gradient Fourier series solution converges more quickly to its limit compared with the Terzaghi one.

Figure 8 shows the comparison between the Terzaghi and second gradient solutions for increasing time. It must be noted that, due to the continuity between the initial datum and the BC, the second gradient solution smoothly decreases with respect to the initial datum, while the Terzaghi solution is not able to describe the behavior of $m_f$ close to the external surface. The second gradient allows for describing compaction of the solid in the vicinity of the external surface, which contrasts instantaneous escape of the fluid out of the porous skeleton. This effect has been indeed recognized both in experiments and in situ measurements [Mollema and Antonellini 1996; Holcomb and Olsson 2003; 2000].

## 6. Concluding remarks

In this paper an application of the second gradient theory of poromechanics to the consolidation problem is discussed. In particular, we present some results within the hypothesis of small deformations around a prestressed reference configuration of the solid skeleton. Even in the framework of the linearized theory, the considered second gradient model gives rise to several interesting questions, concerning both the mathematical formulation of the problem and the mechanical interpretation of the results.

From the mathematical point of view the problem could be studied within the framework of linear pencils of ordinary differential operators on a finite interval, with boundary conditions depending on the spectral parameter. Several applications of this theory to physics and mechanics can be found in the literature [Tretter 2000; Marletta et al. 2003]; it is our purpose to investigate in the future how the very special problem we are dealing with can fit within the general theory.

From the mechanical point of view, the results presented also look quite interesting, in particular concerning the capability of the model to describe fluid segregation. It has to be remarked that second gradient models, in general, regularize the solutions of evolutionary or equilibrium equations (see, for example, Figure 8). In the case of their application to phase transition phenomena they allow for the coexistence of different phases at equilibrium, in the case of strain concentration phenomena for the description of shear and compaction bands, and in the case of wetting for the description of drop/film stability. In the first and third instances, the second gradient is necessary to describe capillarity, and in the second plasticity. In the present instance we propose to use second gradient models to describe those phenomena occurring close to the drained boundary which contrasts fluid flow.

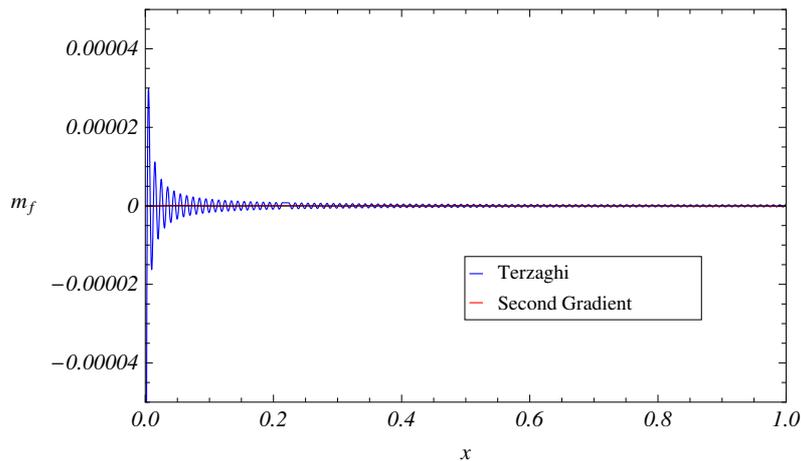

**Figure 7.** Comparison between the classical Terzaghi solution and the second gradient solution for $t = 0^+$.

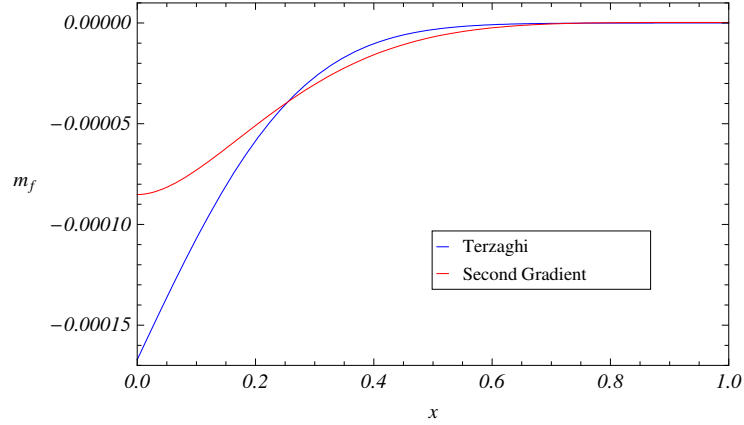

**Figure 8.** Comparison between the Terzaghi and the second gradient solutions for $t = 0.1$.

A linear stability analysis provides the limit value of the initial stress, which forces the first eigenvalues to become at least nonnegative. The goal in the future will be that of identifying which are the buckling modes, and in particular to corroborate the idea that bifurcated modes of consolidation can interpret liquefaction phenomena and occurrence of sand boils [Kolymbas 1998].

### Appendix A: Coefficients of the differential problem

The constant coefficients $C_i$ and $\mathbb{C}_i$ appearing in the differential problem given by Equations (17) and (19) are defined as

$$C_1 = k_1 k_3, \qquad C_2\left(p_0^{\text{ext.}}\right) = k_1 + k_3 k_5 (k_2 + b)^2 + k_3 C_4\left(p_0^{\text{ext.}}\right),$$

$$C_3 = k_4\left(k_1 + k_3 k_5 k_2^2\right), \qquad C_4\left(p_0^{\text{ext.}}\right) = 1 - \frac{p_0^{\text{ext.}}}{\lambda + 2\mu},$$

$$C_5\left(p_0^{\text{ext.}}\right) = k_4\left[C_4\left(p_0^{\text{ext.}}\right) + k_6\right] + k_1 + k_3 k_5 k_2^2, \qquad C_6\left(p_0^{\text{ext.}}\right) = C_4\left(p_0^{\text{ext.}}\right) + k_6,$$

with

$$k_1 = \frac{K_{ss}}{(\lambda + 2\mu)L^2}, \qquad k_2 = K_{sf}, \qquad k_3 = \frac{\mathbb{M}}{ML^2},$$

$$k_4 = \frac{\alpha}{DL^2}, \qquad k_5 = \frac{M}{\lambda + 2\mu}, \qquad k_6 = b^2 k_5.$$

Moreover, the coefficients appearing in the BC, Equation (18), are defined as

$$\mathbb{C}_1 = C_1, \qquad \mathbb{C}_2\left(p_0^{\text{ext.}}\right) = C_2\left(p_0^{\text{ext.}}\right), \qquad \mathbb{C}_3 = C_3$$

$$\mathbb{C}_4\left(p_0^{\text{ext.}}\right) = C_4\left(p_0^{\text{ext.}}\right), \quad \mathbb{C}_5(\pi_0) = C_5 - \left(k_1 + k_3 k_5 k_2^2\right), \quad \mathbb{C}_6\left(\pi_f^0\right) = \frac{\Delta p^{\text{ext.}}}{bM}.$$

It must be noticed that the constants $k_1, \ldots, k_4$ are introduced by the second gradient model, while $k_5$ and $k_6$ are related to the first gradient parameters $M$, $\lambda$, and $\mu$, which represent the Biot bulk modulus and the Lamé coefficients of the considered material, respectively.

ANGELA MADEO: angela.madeo@uniroma1.it
*Dipartimento di Metodi e Modelli Matematici per le Scienze Applicate, Università di Roma "La Sapienza", Via Scarpa 16, 00161 Rome, Italy*

FRANCESCO DELL'ISOLA: francesco.dellisola@uniroma1.it
*Dipartimento di Ingegneria Strutturale e Geotecnica, Università di Roma "La Sapienza", Via Eudossiana 18, 00184 Rome, Italy*

and

*Laboratorio di Strutture e Materiali Intelligenti, Palazzo Caetani (Ala Nord), 04012 Cisterna di Latina, Italy*

NICOLETTA IANIRO: ianiro@dmmm.uniroma1.it
*Dipartimento di Metodi e Modelli Matematici per le Scienze Applicate, Università di Roma "La Sapienza", Via Scarpa 16, 00161 Rome, Italy*

GIULIO SCIARRA: giulio.sciarra@uniroma1.it
*Dipartimento di Ingegneria Chimica Materiali Ambiente, Università di Roma "La Sapienza", Via Eudossiana 18, 00184 Rome, Italy*